\pdfoutput=1
\documentclass[12pt,a4paper]{article}

\usepackage{lineno}  
\usepackage{graphicx}  
\usepackage{xspace} 
\usepackage{color}
\usepackage{colortbl}
\usepackage{amsmath} 
\usepackage{ifthen} 
\graphicspath{{./figs/}} 

\newboolean{pdflatex}
\setboolean{pdflatex}{true} 

\newboolean{articletitles}
\setboolean{articletitles}{true} 

\newboolean{uprightparticles}
\setboolean{uprightparticles}{false} 

\usepackage{amssymb}
\usepackage{amsfonts}
\usepackage{upgreek} 

\usepackage{hyperref}    
\usepackage[all]{hypcap} 

\def\lhcb {\mbox{LHCb}\xspace}
\def\ux85 {\mbox{UX85}\xspace}

\def\babar  {\mbox{BaBar}\xspace}

\ifthenelse{\boolean{uprightparticles}}%
{

 \def\PDelta      {\ensuremath{\Delta}\xspace}                 
 \def\PXi      {\ensuremath{\Xi}\xspace}                 
 \def\PLambda      {\ensuremath{\Lambda}\xspace}                 
 \def\PSigma      {\ensuremath{\Sigma}\xspace}                 
 \def\POmega      {\ensuremath{\Omega}\xspace}                 
 \def\PUpsilon      {\ensuremath{\Upsilon}\xspace}                 
 
 \def\PB      {\ensuremath{\mathrm{B}}\xspace}                 
                  
 \def\PD      {\ensuremath{\mathrm{D}}\xspace}

 \def\PK      {\ensuremath{\mathrm{K}}\xspace}

 \def\Pb      {\ensuremath{\mathrm{b}}\xspace}                 
 \def\Pc      {\ensuremath{\mathrm{c}}\xspace}

 \def\Pi      {\ensuremath{\mathrm{i}}\xspace}

 \def\Ps      {\ensuremath{\mathrm{s}}\xspace}

}
{

 \mathchardef\PDelta="7101
 \mathchardef\PXi="7104
 \mathchardef\PLambda="7103
 \mathchardef\PSigma="7106
 \mathchardef\POmega="710A
 \mathchardef\PUpsilon="7107
                  
 \def\PB      {\ensuremath{B}\xspace}                 
                  
 \def\PD      {\ensuremath{D}\xspace}

 \def\PK      {\ensuremath{K}\xspace}

 \def\Pb      {\ensuremath{b}\xspace}                 
 \def\Pc      {\ensuremath{c}\xspace}

 \def\Pi      {\ensuremath{i}\xspace}

 \def\Ps      {\ensuremath{s}\xspace}

}

\def\squark    {\ensuremath{\Ps}\xspace}
\def\squarkbar {\ensuremath{\overline \squark}\xspace}

\def\cquark    {\ensuremath{\Pc}\xspace}

\def\bquark    {\ensuremath{\Pb}\xspace}

\def\kaon  {\ensuremath{\PK}\xspace}
  \def\Kbar  {\kern 0.2em\overline{\kern -0.2em \PK}{}\xspace}

\def\Kz    {\ensuremath{\kaon^0}\xspace}
\def\Kzb   {\ensuremath{\Kbar^0}\xspace}
\def\KzKzb {\ensuremath{\Kz \kern -0.16em \Kzb}\xspace}
\def\Kp    {\ensuremath{\kaon^+}\xspace}
\def\Km    {\ensuremath{\kaon^-}\xspace}

\def\KpKm  {\ensuremath{\Kp \kern -0.16em \Km}\xspace}
\def\KS    {\ensuremath{\kaon^0_{\rm\scriptscriptstyle S}}\xspace}

  \def\Dbar    {\kern 0.2em\overline{\kern -0.2em \PD}{}\xspace}
\def\D       {\ensuremath{\PD}\xspace}

\def\Dz      {\ensuremath{\D^0}\xspace}
\def\Dzb     {\ensuremath{\Dbar^0}\xspace}
\def\DzDzb   {\ensuremath{\Dz {\kern -0.16em \Dzb}}\xspace}
\def\Dp      {\ensuremath{\D^+}\xspace}
\def\Dm      {\ensuremath{\D^-}\xspace}

\def\DpDm    {\ensuremath{\Dp {\kern -0.16em \Dm}}\xspace}

  \def\Bbar    {\kern 0.18em\overline{\kern -0.18em \PB}{}\xspace}

  \def\Y#1S{\ensuremath{\PUpsilon{(#1S)}}\xspace}

\def\Lbar {\ensuremath{\kern 0.1em\overline{\kern -0.1em\Lambda\kern -0.05em}\kern 0.05em{}}\xspace}

\def\BF         {{\ensuremath{\cal B}\xspace}}

\def\to                 {\ensuremath{\rightarrow}\xspace}

\def\AT#1     {\ensuremath{A_{\mathrm{T}}^{#1}}\xspace}           

\def\C#1      {\ensuremath{\mathcal{C}_{#1}}\xspace}                       
\def\Cp#1     {\ensuremath{\mathcal{C}_{#1}^{'}}\xspace}                    
\def\Ceff#1   {\ensuremath{\mathcal{C}_{#1}^{\mathrm{(eff)}}}\xspace}        
\def\Cpeff#1  {\ensuremath{\mathcal{C}_{#1}^{'\mathrm{(eff)}}}\xspace}       
\def\Ope#1    {\ensuremath{\mathcal{O}_{#1}}\xspace}                       
\def\Opep#1   {\ensuremath{\mathcal{O}_{#1}^{'}}\xspace}                    

\newcommand{\tev}{\ensuremath{\mathrm{\,Te\kern -0.1em V}}\xspace}
\newcommand{\gev}{\ensuremath{\mathrm{\,Ge\kern -0.1em V}}\xspace}
\newcommand{\mev}{\ensuremath{\mathrm{\,Me\kern -0.1em V}}\xspace}
\newcommand{\kev}{\ensuremath{\mathrm{\,ke\kern -0.1em V}}\xspace}
\newcommand{\ev}{\ensuremath{\mathrm{\,e\kern -0.1em V}}\xspace}
\newcommand{\gevc}{\ensuremath{{\mathrm{\,Ge\kern -0.1em V\!/}c}}\xspace}
\newcommand{\mevc}{\ensuremath{{\mathrm{\,Me\kern -0.1em V\!/}c}}\xspace}
\newcommand{\gevcc}{\ensuremath{{\mathrm{\,Ge\kern -0.1em V\!/}c^2}}\xspace}
\newcommand{\gevgevcccc}{\ensuremath{{\mathrm{\,Ge\kern -0.1em V^2\!/}c^4}}\xspace}
\newcommand{\mevcc}{\ensuremath{{\mathrm{\,Me\kern -0.1em V\!/}c^2}}\xspace}

\def\mm   {\ensuremath{\rm \,mm}\xspace}

\def\mum  {\ensuremath{\,\upmu\rm m}\xspace}

\def\invfb   {\ensuremath{\mbox{\,fb}^{-1}}\xspace}

\def\gsim{{~\raise.15em\hbox{$>$}\kern-.85em
          \lower.35em\hbox{$\sim$}~}\xspace}
\def\lsim{{~\raise.15em\hbox{$<$}\kern-.85em
          \lower.35em\hbox{$\sim$}~}\xspace}

\def\pt         {\mbox{$p_{\rm T}$}\xspace}

\def\evtgen     {\mbox{\textsc{EvtGen}}\xspace}
\def\pythia     {\mbox{\textsc{Pythia}}\xspace}

\def\geant      {\mbox{\textsc{Geant4}}\xspace}

\def\tell1  {TELL1\xspace}
\def\ukl1   {UKL1\xspace}

\usepackage{cite} 
\usepackage{mciteplus}

\begin{document}

\renewcommand{\thefootnote}{\fnsymbol{footnote}}
\setcounter{footnote}{1}

\begin{titlepage}
\pagenumbering{roman}

\vspace*{-1.5cm}
\centerline{\large EUROPEAN ORGANIZATION FOR NUCLEAR RESEARCH (CERN)}
\vspace*{1.5cm}
\hspace*{-0.5cm}
\begin{tabular*}{\linewidth}{lc@{\extracolsep{\fill}}r}
\ifthenelse{\boolean{pdflatex}}
{\vspace*{-2.7cm}\mbox{\!\!\!\includegraphics[width=.14\textwidth]{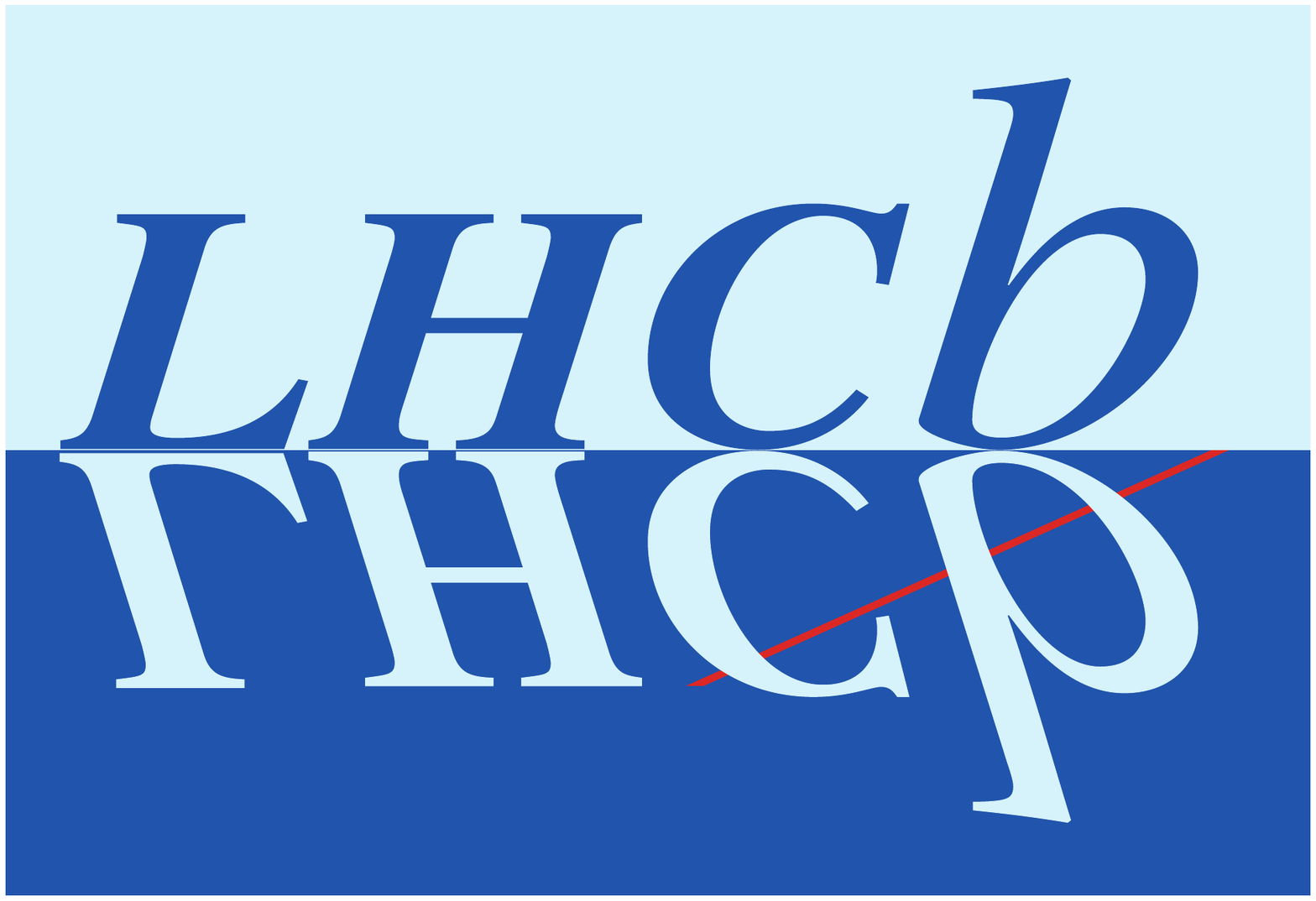}} & &}%
{\vspace*{-1.2cm}\mbox{\!\!\!\includegraphics[width=.12\textwidth]{figs/lhcb-logo.eps}} & &}%
\\
 & & CERN-PH-EP-2012-181 \\  
 & & LHCb-PAPER-2012-016 \\  
 & & 24 July 2012 \\
 & & \\
\end{tabular*}

\vspace*{4.0cm}

{\bf\boldmath\huge
\begin{center}
Study of $D_{sJ}$ decays to $D^+\KS$ and $D^0K^+$ final states in $pp$ collisions
\end{center}
}

\vspace*{2.0cm}

\begin{center}
The LHCb collaboration\footnote{Authors are listed on the following pages.}
\end{center}

\vspace{\fill}

\begin{abstract}
  \noindent
A study of $D^+\KS$ and $D^0K^+$ final states is performed in a sample of 1.0\invfb of $pp$ collision data collected at a centre-of-mass energy of $\sqrt{s}=7\tev$ with the \lhcb detector. We confirm the existence of the $D_{s1}^*(2700)^+$ and $D_{sJ}^*(2860)^+$ excited states and measure their masses and widths to be
\begin{eqnarray}
m(D_{s1}^*(2700)^+) &=& 2709.2 \pm 1.9(\mbox{stat})\pm\,\,\,4.5(\mbox{syst})~\mevcc,\cr
\Gamma(D_{s1}^*(2700)^+) &=& \,\,\,115.8 \pm 7.3(\mbox{stat}) \pm12.1(\mbox{syst})~\mevcc,\cr
m(D_{sJ}^*(2860)^+) &=& 2866.1 \pm 1.0(\mbox{stat}) \pm\,\,\,6.3(\mbox{syst})~\mevcc,\cr
\Gamma(D_{sJ}^*(2860)^+) &=& \,\,\,\,\,\,69.9 \pm 3.2(\mbox{stat}) \pm\,\,\,6.6(\mbox{syst})~\mevcc.\cr\nonumber
\end{eqnarray}

\end{abstract}

\vspace*{0.5cm}

\begin{center}
  {\it Submitted to JHEP}
\end{center}

\vspace{\fill}

\end{titlepage}

\newpage
\setcounter{page}{2}
\mbox{~}
\newpage

\centerline{\large\bf LHCb collaboration}
\begin{flushleft}
\small
R.~Aaij$^{38}$, 
C.~Abellan~Beteta$^{33,n}$, 
A.~Adametz$^{11}$, 
B.~Adeva$^{34}$, 
M.~Adinolfi$^{43}$, 
C.~Adrover$^{6}$, 
A.~Affolder$^{49}$, 
Z.~Ajaltouni$^{5}$, 
J.~Albrecht$^{35}$, 
F.~Alessio$^{35}$
% ...

{\footnotesize \it
% ...
$ ^{5}$Clermont Universit\'{e}, Universit\'{e} Blaise Pascal, CNRS/IN2P3, LPC, Clermont-Ferrand, France\\
$ ^{6}$CPPM, Aix-Marseille Universit\'{e}, CNRS/IN2P3, Marseille, France\\
% ...
$ ^{11}$Physikalisches Institut, Ruprecht-Karls-Universit\"{a}t Heidelberg, Heidelberg, Germany\\
% ...
$ ^{33}$Universitat de Barcelona, Barcelona, Spain\\
$ ^{34}$Universidad de Santiago de Compostela, Santiago de Compostela, Spain\\
$ ^{35}$European Organization for Nuclear Research (CERN), Geneva, Switzerland\\
% ...
$ ^{38}$Nikhef National Institute for Subatomic Physics, Amsterdam, The Netherlands\\
% ...
$ ^{43}$H.H. Wills Physics Laboratory, University of Bristol, Bristol, United Kingdom\\
% ...
$ ^{49}$Oliver Lodge Laboratory, University of Liverpool, Liverpool, United Kingdom\\
% ...
$ ^{n}$LIFAELS, La Salle, Universitat Ramon Llull, Barcelona, Spain\\
% ...
}
\end{flushleft}

\cleardoublepage

\renewcommand{\thefootnote}{\arabic{footnote}}
\setcounter{footnote}{0}

\pagestyle{plain} 
\setcounter{page}{1}
\pagenumbering{arabic}


\section{Introduction}
\label{sec:Introduction}
The spectrum of the known $\cquark\squarkbar$ states is at present described as two S-wave states ($D_{s}^+$, $D_s^{*+}$) with spin-parity assignment $J^P=0^-,\,1^-$ and four P-wave states ($D_{s0}^*(2317)^+$, $D_{s1}(2460)^+$, $D_{s1}(2536)^+$, $D_{s2}^*(2573)^+$) with $J^P=0^+,\,1^+,\,1^+,\,2^+$~\cite{PDG2012}, of which the latter two have also been observed in semileptonic $B$-decays in LHCb~\cite{Aaij:2011ju}. This picture is still controversial since the $D_{s0}^*(2317)^+$ and $D_{s1}(2460)^+$ states, discovered in 2003~\cite{Aubert:2003fg,Besson:2003cp,Abe:2003jk,Aubert:2003pe}, were predicted to have much higher masses~\cite{Godfrey:1985xj,Godfrey:1986wj,Isgur:1991wq,DiPierro:2001uu,Matsuki:2007zza}. 
Between 2006 and 2009, three new $D_{sJ}$ mesons were observed at the $B$ factories in $DK$ and $D^*K$ decay modes\footnote{$DK$ refers to $D^+\KS$ and $D^0K^+$, while $D^*K$ refers to $D^{*+}\KS$ and $D^{*0}K^+$ final states, where the inclusion of charge conjugate final states is implicit everywhere.} and in three-body $b$-hadron decays: the $D_{s1}^*(2700)^+$~\cite{Aubert:2006mh,Brodzicka:2007aa,Aubert:2009ah}, the $D_{sJ}^*(2860)^+$~\cite{Aubert:2006mh,Aubert:2009ah} and the $D_{sJ}(3040)^+$~\cite{Aubert:2009ah} excited states. From the angular analyses in Refs.~\cite{Aubert:2009ah,Brodzicka:2007aa}, $J^P=1^-$ is favoured for the $D_{s1}^*(2700)^+$ state, a possible $J^P=3^-$ assignment is discussed for the $D_{sJ}^*(2860)^+$, and an unnatural parity is suggested for the $D_{sJ}(3040)^+$ state since it was found to decay only to the $D^*K$ final state. 

The measured properties of the $D_{s1}^*(2700)^+$ state are in agreement with theoretical expectations~\cite{Godfrey:1985xj,Godfrey:1986wj,Isgur:1991wq,DiPierro:2001uu}, but further confirmation 
is still needed. Similarly, the existence of the $D_{sJ}^*(2860)^+$ resonance is unclear. In the latest analysis by the \babar collaboration~\cite{Aubert:2009ah}, the observation of the $D_{sJ}^*(2860)^+$ decaying to the $D^{*}K$ final state rules out the $J^P=0^+$ assignment of Refs.~\cite{vanBeveren:2006st,Close:2006gr}, reinforcing the $3^-$ assignment proposed in Refs.~\cite{Colangelo:2006rq, Zhang:2006yj}. However, the measured branching fraction ratio $\BF(D_{sJ}^*(2860)^+\to D^{*}K)/\BF(D_{sJ}^*(2860)^+\to DK)=1.1\pm0.2$ is larger than the predicted values of 0.39~\cite{Colangelo:2006rq} and 0.6~\cite{Zhang:2006yj}. 
In Ref.~\cite{vanBeveren:2009jq}, it is argued that the observed decays of the $D_{sJ}^{*}(2860)^+$ to $DK$ and $D^{*}K$ do not rule out a $J^P=0^+$ assignment, but the pattern of decays can be explained by the overlap of two radially excited states of $J^P=0^+$ and $J^P=2^+$, with similar mass and width.

Given the controversial status of these high mass $D_{sJ}$ states, none of them is currently reported in the summary table of the Particle Data Group~\cite{PDG2012}. Experimental contributions are needed in order to disentangle the puzzle around the $D_{sJ}^*(2860)^+$ and to complete the picture of the $\cquark\squarkbar$ spectrum.

Using 1.0\invfb of data recorded by the \lhcb detector during 2011 we perform an analysis of the $D^+\KS$ and $D^0K^+$ final states

in order to confirm the existence of the $D_{s1}^*(2700)^+$ and $D_{sJ}^*(2860)^+$ states and to measure their masses and widths. 

\section{Detector description}
\label{sec:Detector}

The \lhcb detector~\cite{Alves:2008zz} is a single-arm forward
spectrometer covering the \mbox{pseudorapidity} range $2<\eta <5$, designed
for the study of particles containing \bquark or \cquark quarks. The
detector includes a high precision tracking system consisting of a
silicon-strip vertex detector surrounding the $pp$ interaction region,
a large-area silicon-strip detector located upstream of a dipole
magnet with a bending power of about $4{\rm\,Tm}$, and three stations
of silicon-strip detectors and straw drift-tubes placed
downstream. The combined tracking system has momentum resolution
$\Delta p/p$ that varies from 0.4\% at 5\gevc to 0.6\% at 100\gevc,
and impact parameter\footnote{The perpendicular distance between the track path and the position of a $pp$ collision.} resolution of 20\mum for tracks with high
transverse momentum (\pt) with respect to the beam direction. Charged hadrons are identified using two
ring-imaging Cherenkov detectors. Photon, electron and hadron
candidates are identified by a calorimeter system consisting of
scintillating-pad and pre-shower detectors, an electromagnetic
calorimeter and a hadronic calorimeter. Muons are identified by a muon
system composed of alternating layers of iron and multiwire
proportional chambers. The trigger consists of a hardware stage, based
on information from the calorimeter and muon systems, followed by a
software stage which applies a full event reconstruction. 

Monte Carlo simulated event samples are used to calculate the effects of the detector on the mass resolution. The $pp$ collisions are generated using
\pythia~6.4~\cite{Sjostrand:2006za} with a specific \lhcb
configuration~\cite{LHCb-PROC-2010-056}. Decays of hadronic particles
are described by \evtgen~\cite{Lange:2001uf} and the
interaction of the generated particles with the detector and its
response are implemented using the \geant
toolkit~\cite{Allison:2006ve,Agostinelli:2002hh} as described in
Ref.~\cite{LHCb-PROC-2011-006}. Simulated events are reconstructed in the same manner as data.

\section{Event selection}
\label{sec:Selection}
\begin{figure}[tb]
  \begin{center}
    \includegraphics[width=0.49\textwidth]{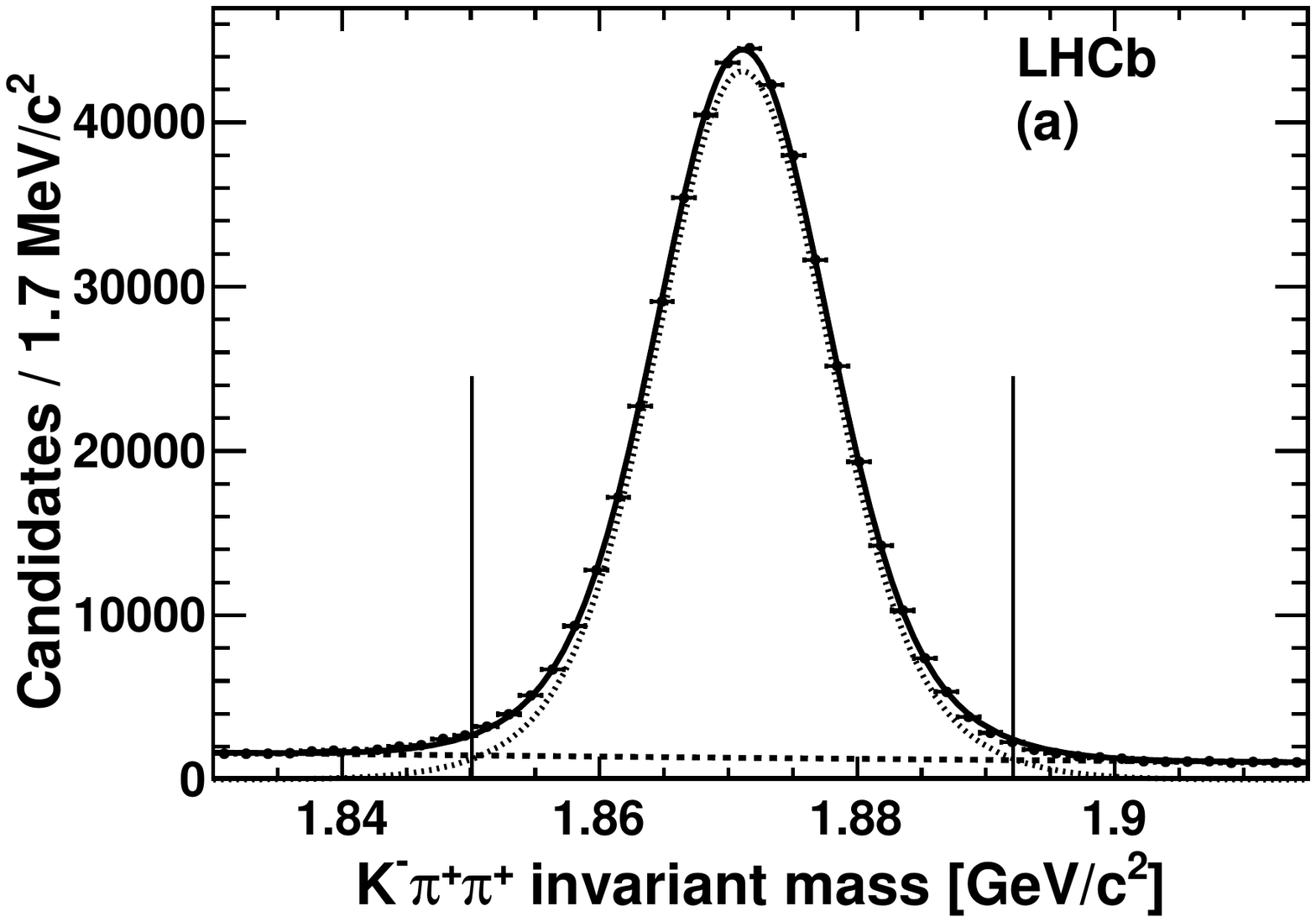}
    \includegraphics[width=0.49\textwidth]{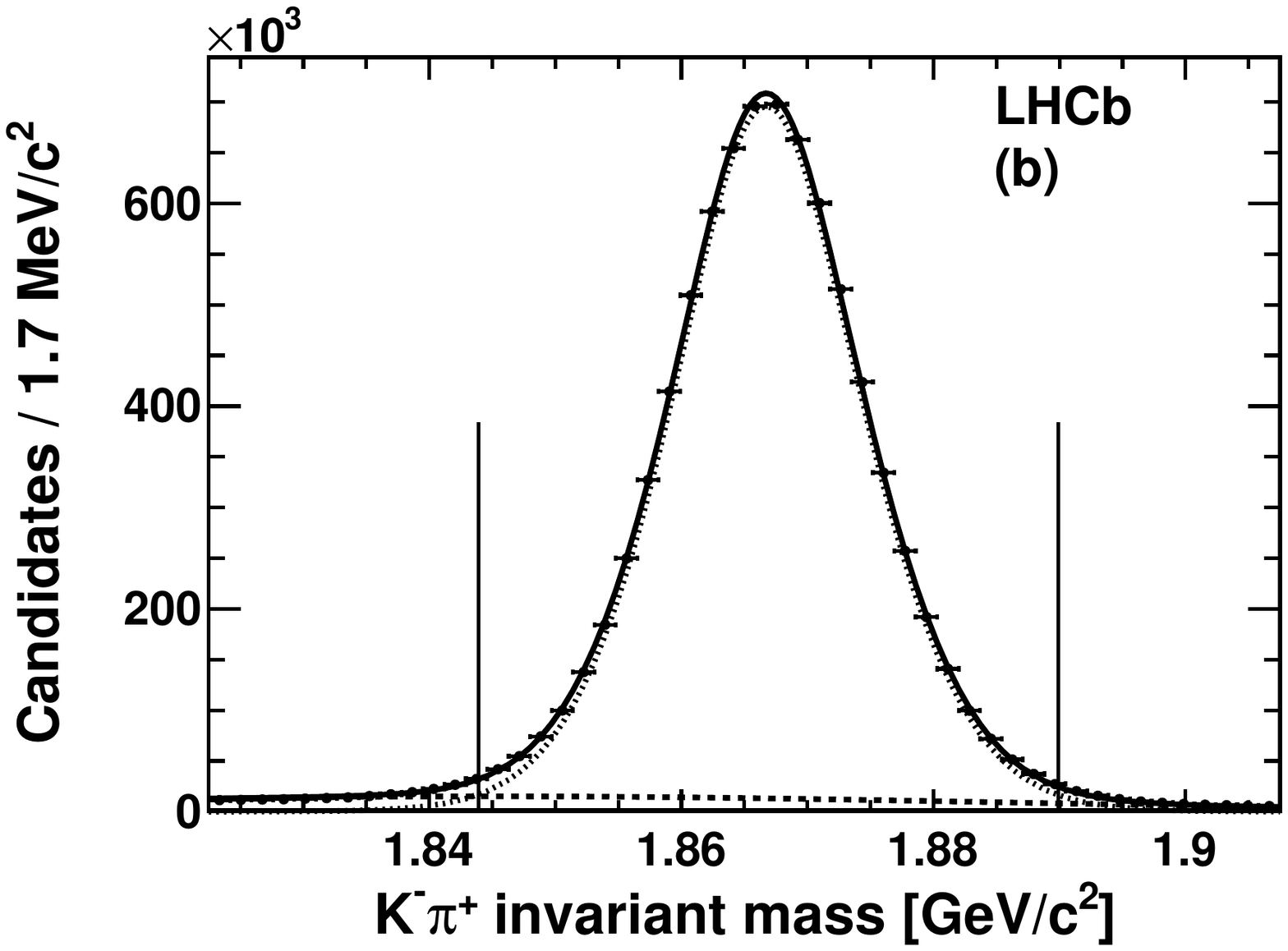}
    \includegraphics[width=0.49\textwidth]{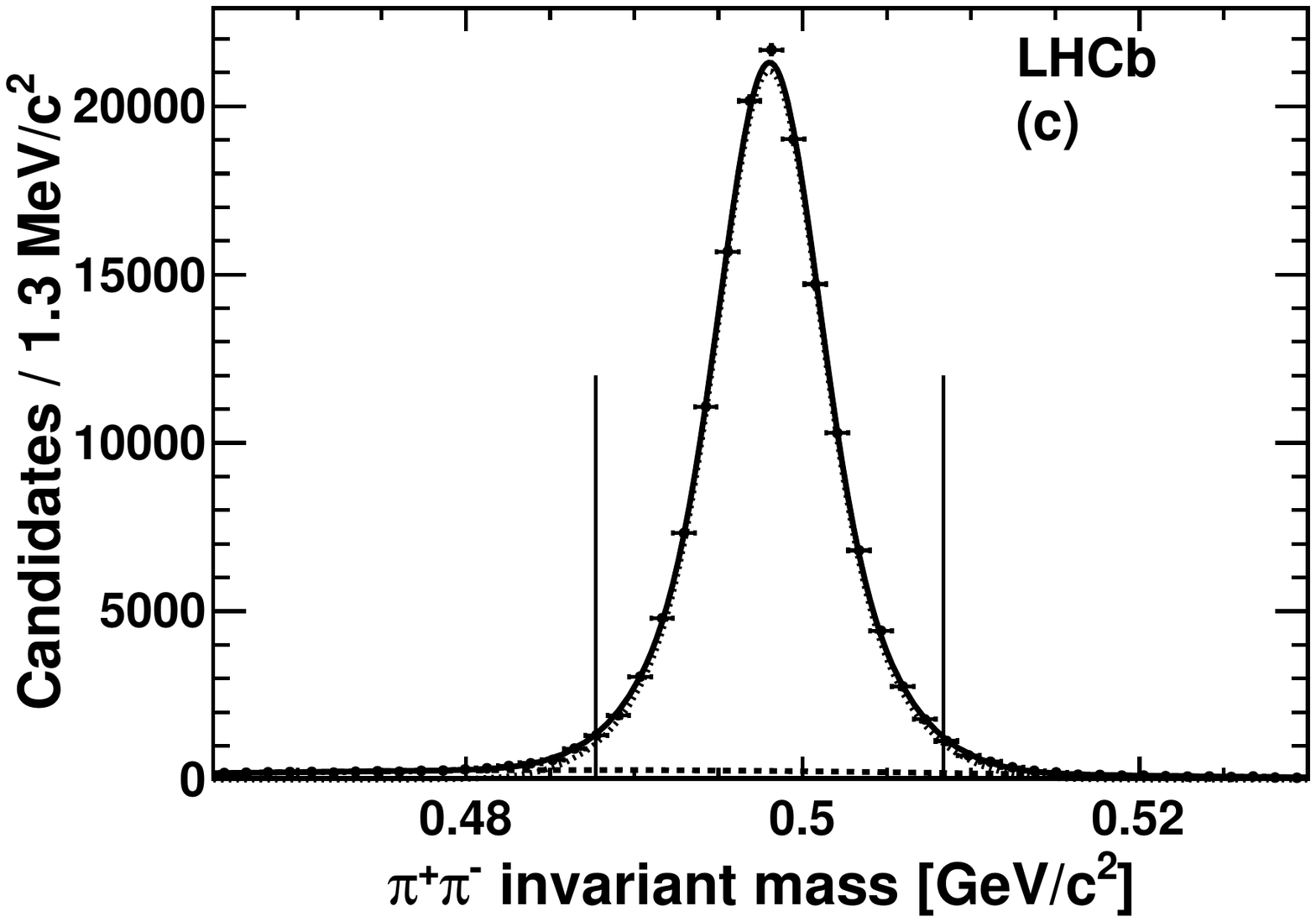}
    \includegraphics[width=0.49\textwidth]{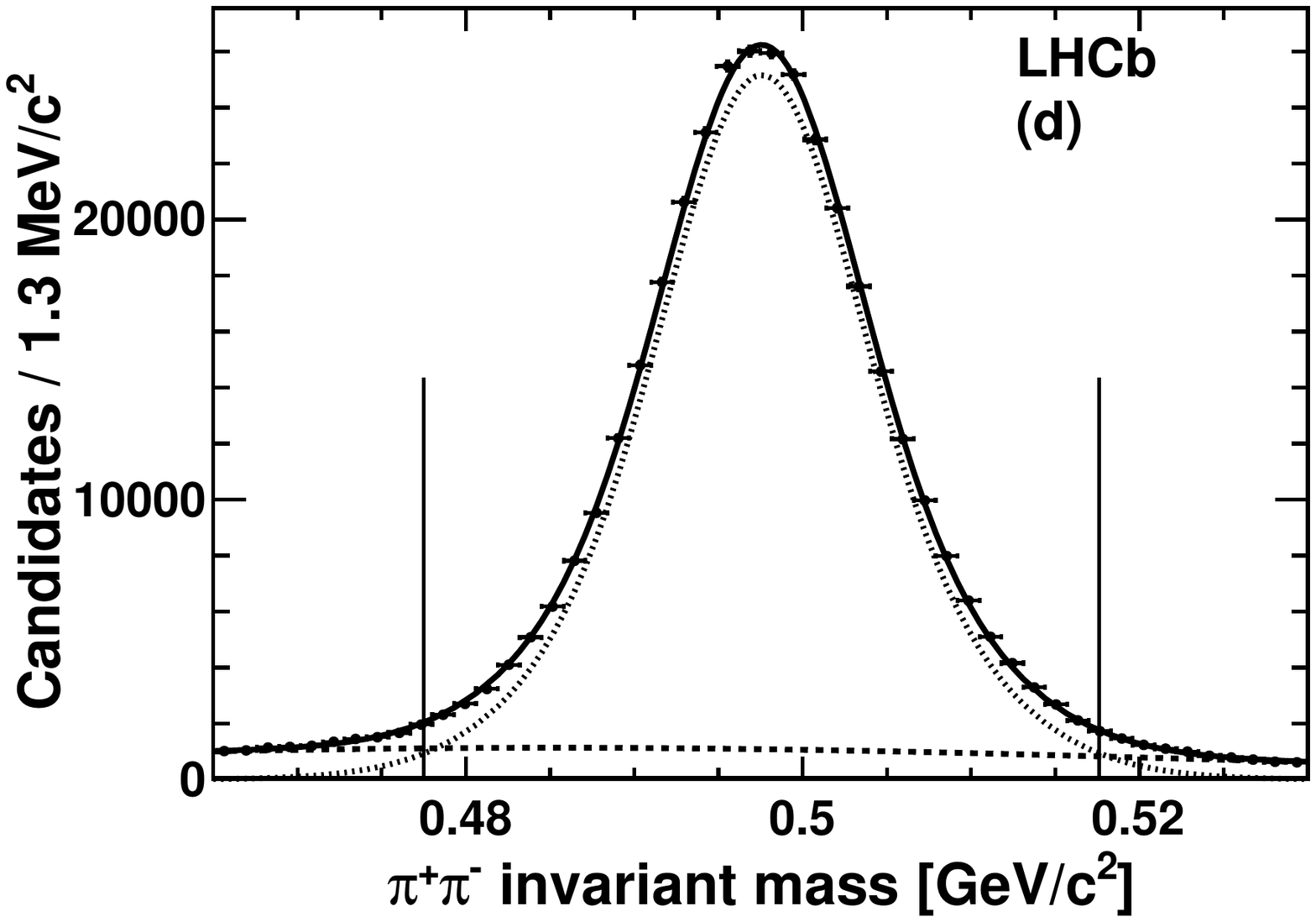}
    \vspace*{-0.5cm}
  \end{center}
  \caption{
    \small Invariant mass distribution (points) for (a) $D^+$, (b) $D^0$, $\KS$ decaying (c) inside and (d) outside the vertex detector. We show the total probability density function (solid curve), the signal component as a sum of Gaussian distributions (dotted curve) and a decreasing exponential distribution to describe the background component (dashed curve). The region within the vertical lines corresponds to $\pm3$ standard deviations in mass resolution from the measured peak.
    }
  \label{fig:massfits}
\end{figure}

We reconstruct the $D^+\KS$ final state using the $D^+\to K^-\pi^+\pi^+$ and $\KS\to\pi^+\pi^-$ decay modes, and the $D^0K^+$ final state using the $D^0\to K^-\pi^+$ decay mode. Because of their long lifetime, \KS mesons may decay inside or outside the vertex detector. Those that decay within the vertex detector acceptance have a mass resolution about half as large as those that decay outside of its acceptance, as observed in Fig.~\ref{fig:massfits}.

Tracks are required to have good track fit quality, momentum $p>3\gevc$ and transverse momentum $\pt>250\mevc$. Tracks pointing to a $pp$ collision vertex (primary vertex) are rejected by means of an impact parameter requirement in the reconstruction of the $D^+$, $D^0$ and $\KS$ candidates. 
The tracks used to reconstruct the mesons decaying inside the vertex detector are required to have a distance of closest approach among them smaller than 0.5\mm.
 
To improve the signal to background ratio for the reconstructed $D^+$, $D^0$ and $\KS$ meson candidates, we require the cosine of the angle between the momentum of the meson candidate and the direction defined by the positions of the primary and the meson decay vertex, to be larger than 0.9999 for $\KS$ and 0.99999 for charmed mesons. This requirement ensures that the meson candidates are produced in the primary $pp$ interaction, and reduces the contribution from particles originating from $b$-hadron decays. The $D^+$ and $\KS$, and similarly $D^0$ and $K^+$ candidates, are fitted to a common vertex requiring $\chi^2/\mbox{ndf}<8$, where ndf is the number of degrees of freedom. 
The purity of the charmed meson candidates is enhanced by requiring the decay products to be identified by the ring-imaging Cherenkov detectors, using the difference in the log-likelihood between the kaon and pion hypotheses $\Delta\ln\mathcal{L}_{K\pi}$. We require $\Delta\ln\mathcal{L}_{K\pi}>2(0)$ for kaon tracks and $\Delta\ln\mathcal{L}_{K\pi}<10(6)$ for pion tracks from $D^+(D^0)$ decays. 
The overlap region in the particle identification definition of a kaon and a pion is small and not a problem given the reduced number of multiple candidates per event.
Figure~\ref{fig:massfits} shows the invariant mass spectra for the $D^+$, $D^0$ and $\KS$ meson candidates after the described selection is applied. The signal regions for $D^+$, $D^0$ and \KS candidates correspond to $\pm$3 standard deviations in mass resolution from the peak values. 

At $7\tev$, charged track multiplicities from $pp$ interactions are very high, extending beyond 100 tracks per event, leading to large combinatorial background.  We define $\theta$ as the angle between the momentum direction of the kaon in the $DK$ rest frame and the momentum direction of the $DK$ system in the laboratory frame. This variable is symmetrically distributed around zero for resonant states, but more than 90\% of combinatorial background events are in the negative $\cos\theta$ region. We therefore require $\cos\theta>0$ to strongly reduce combinatorial background, for both $D^+\KS$ and $D^0K^+$ final states.
A further reduction of this type of background is achieved by performing an optimization of the signal significance of the cleanest $D_{sJ}$ peak in the $DK$ samples, the $D_{s2}^*(2573)^+$ state. In the $2.5-2.6$\gevcc mass region of the $DK$ spectra, we compute the maximum of the signal significance $N_S/\sqrt{N_S+N_B}$, where $N_S$ and $N_B$ are the number of signal and background events, as a function of different requirements on discriminating variables. This study motivates the following choices. For the $D^+\KS$ final state we require $\pt(D^+\KS)>4.5$\gevc for $\KS$ candidates decaying inside the vertex detector, and $\pt(\KS)>1.5$\gevc for $\KS$ candidates decaying outside the vertex detector. For the $D^0K^+$ final state we require $\pt(K^+)>1.5$\gevc and $P_{\text{NN}_K}(K^+)>0.45$, 
trained using inclusive fully simulated Monte Carlo samples and calculated from a neural network using as input particle identification log-likelihoods, momenta, tracking related variables and sub-detector acceptance requirements combined with Bayesian statistical methods~\cite{Feindt:2006pm}.

\section{ Analysis of the $\boldsymbol{DK}$ invariant mass spectra}
\label{sec:Fit}

\begin{figure}[tb]
  \begin{center}
  \includegraphics[width=0.49\textwidth]{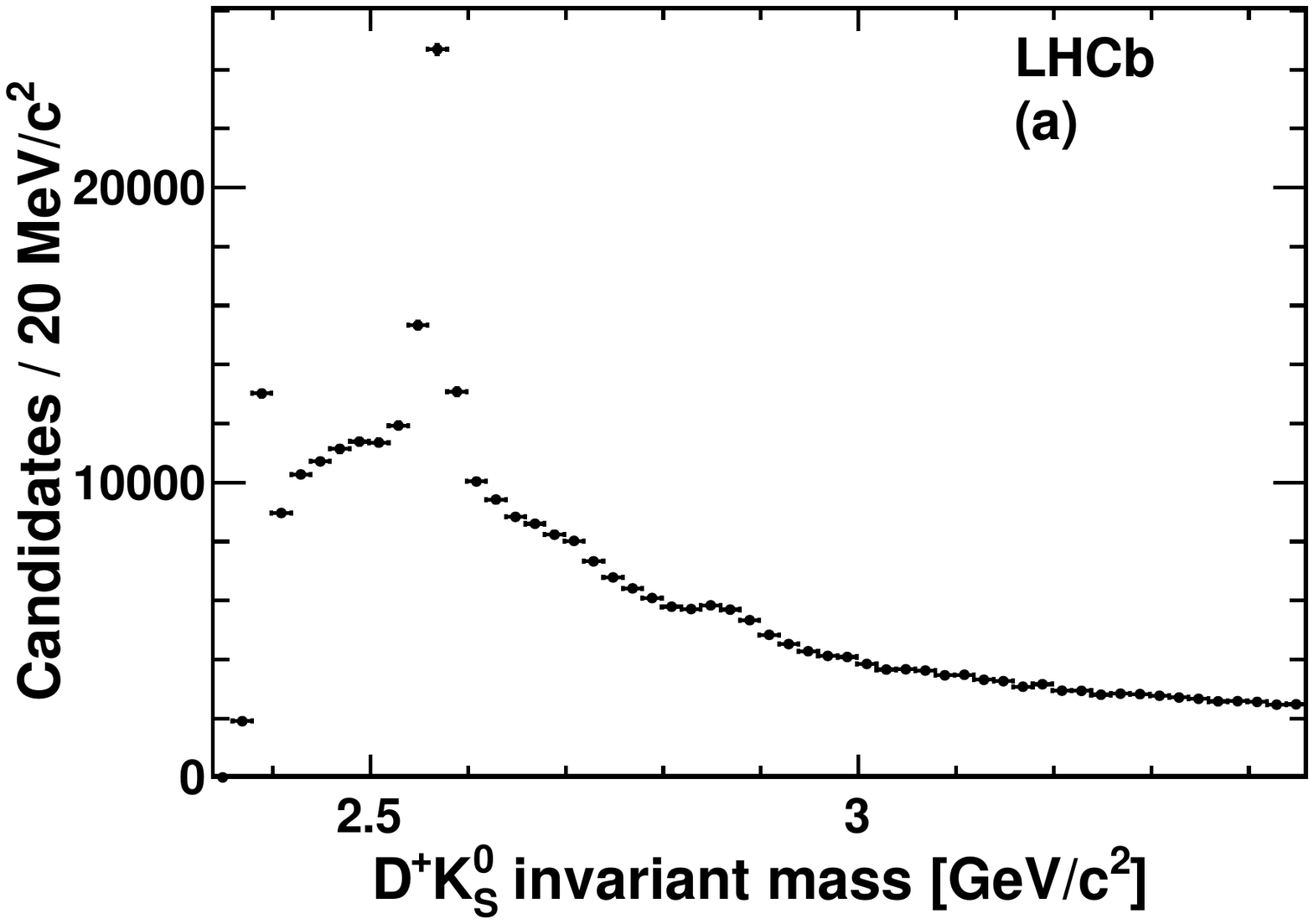}
  \includegraphics[width=0.49\textwidth]{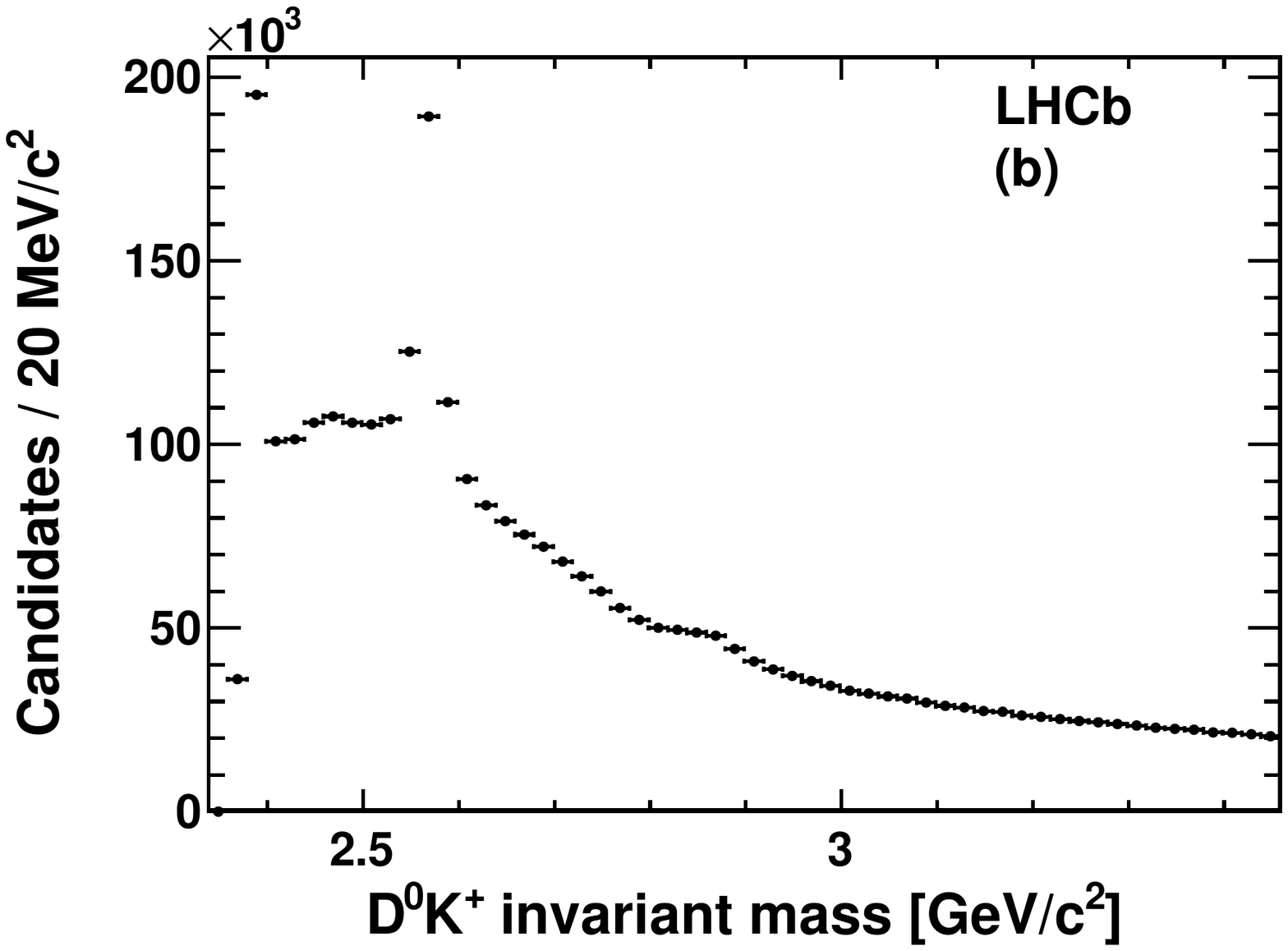}
    \vspace*{-0.5cm}
  \end{center}
  \caption{
    \small Invariant mass distributions for (a) $D^+\KS$ and (b) $D^0K^+$. 
    }
  \label{fig:invmasses}
\end{figure}

The resulting $D^+\KS$ and $D^0K^+$ invariant mass distributions are shown in Fig.~\ref{fig:invmasses}, where we have reconstructed about $0.36\times10^6$ $D^+\KS$ and $3.15\times10^6$ $D^0K^+$ candidates with a multiplicity of 1.005 and 1.010 candidates per event. The $D^+\KS$ and $D^0K^+$ mass spectra show very similar features. The sharp peak near the threshold is due to the feed-down from $D_{s1}(2536)^+\to D^{*+}\KS,\,D^{*0}K^+$ decays, with $D^{*+}\to D^+\pi^0,\,D^+\gamma$ and $D^{*0}\to D^0\pi^0,\,D^0\gamma$, where the neutral pion or photon have not been reconstructed. Since the $D_{s1}(2536)^+$ state has $J^P=1^+$, the decay to $DK$ systems is forbidden by angular momentum and parity conservation. The observed feed-down is well isolated and the overlap with high mass structures is negligible. A prominent peak is observed around 2.57\gevcc, corresponding to the spin-2 $D_{s2}^*(2573)^+$ resonance. We also observe two broad structures near 2.71\gevcc and 2.86\gevcc in both mass spectra, which previous measurements~\cite{Aubert:2009ah} have associated with the spin-1 $D_{s1}^*(2700)^+$ state and the $D_{sJ}^*(2860)^+$ state.

We perform a binned (5\mevcc bin size) simultaneous extended maximum likelihood fit to the two $DK$ mass spectra in the $2.44-3.46$\gevcc range, where the lower bound excludes the $D_{s1}(2536)^+$ feed-down events. Hereafter we will refer to this as the reference fit.

The $D_{sJ}$ signal components are described by relativistic Breit-Wigner lineshapes including the Blatt-Weisskopf form factors which limit the maximum angular momentum in a strong decay via the introduction of an effective radial meson potential~\cite{BW}. Mass resolution effects are neglected in the reference fit, since the expected widths for the $D_{s1}^*(2700)^+$ and $D_{sJ}^*(2860)^+$ states are between one and two orders of magnitude larger than the detector mass resolution, but these effects are included as a source of systematic uncertainty. The background distribution is largely dominated by randomly associated $DK$ pairs created during the hadronization processes, and is described using a linear combination of Chebyshev polynomials of the first kind, of order from one to six. 
These polynomials are flexible and capable of describing possible background fluctuations from non-resonant events. 
The analytical function to describe the background component was trained on a fully combinatorial wrong-sign sample of $D^0K^-$ events, reconstructed and selected in the same way as the $D^0K^+$ final state candidates. 
Additionally, we generate a sample of signal events where the $D_{sJ}$ components of the probability density function are taken from the combined $DK$ and $D^*K$ measurement performed by the \babar experiment~\cite{Aubert:2009ah}. From the combination of the wrong-sign and signal simulated samples we study possible fit instabilities and correlations of the width of the $D_{s1}^*(2700)^+$ state as a function of the lower fit bound.

The signal model was chosen from a set of fits to the $DK$ mass spectra, where we include and remove the expected $D_{sJ}$ states from the fit function, with their masses and widths fixed to the previous \babar measurement. The reference signal model, which shows the best $\chi^2$/ndf, includes the spin-2 $D_{s2}^*(2573)^+$, spin-1 $D_{s1}^*(2700)^+$ and $D_{sJ}^*(2860)^+$ states.
Regarding the $D_{sJ}(2860)^+$ state, we use a spin-0 hypothesis since at present no conclusive $J^P$ assignment has been made for this state. With the current data sample we are not able to identify the presence of additional states in the 2.86\gevcc region, as proposed in Ref.~\cite{vanBeveren:2009jq}. In order to reduce correlations between the background function and the width of the broad resonances and to improve fit stability, we fix the less contributing and most correlated parameters, the order three, five, and six Chebyshev polynomial coefficients for the two $DK$ invariant mass spectra. These parameters are taken from a preliminary fit, where the signal model is fixed to values obtained using an approximate background shape, similar to that used in the \babar analysis~\cite{Aubert:2009ah} and described in Section~\ref{sec:Systematics}. 

\begin{figure}[tb]
  \begin{center}
  \includegraphics[width=0.49\textwidth]{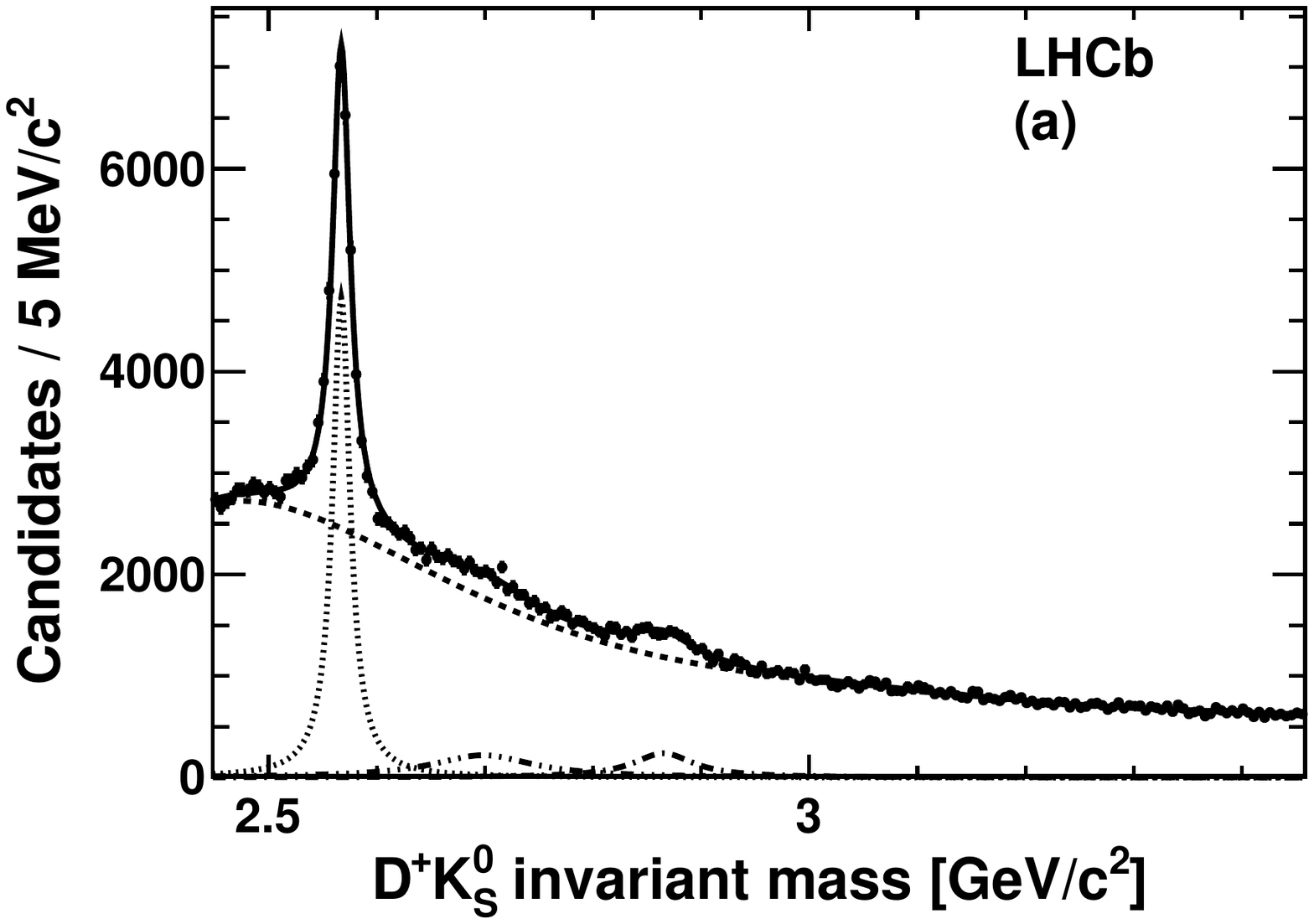}
  \includegraphics[width=0.49\textwidth]{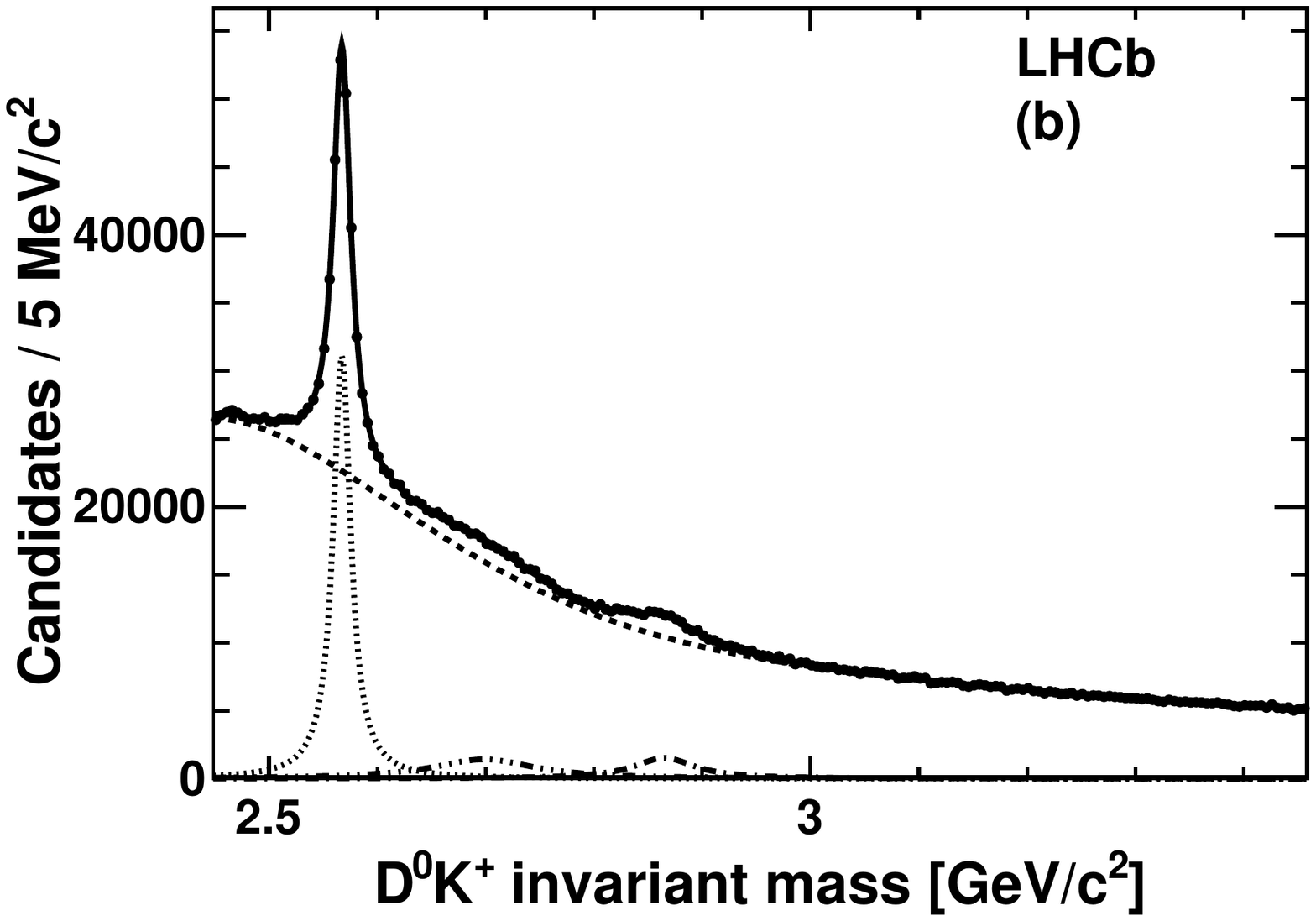}
  \includegraphics[width=0.49\textwidth]{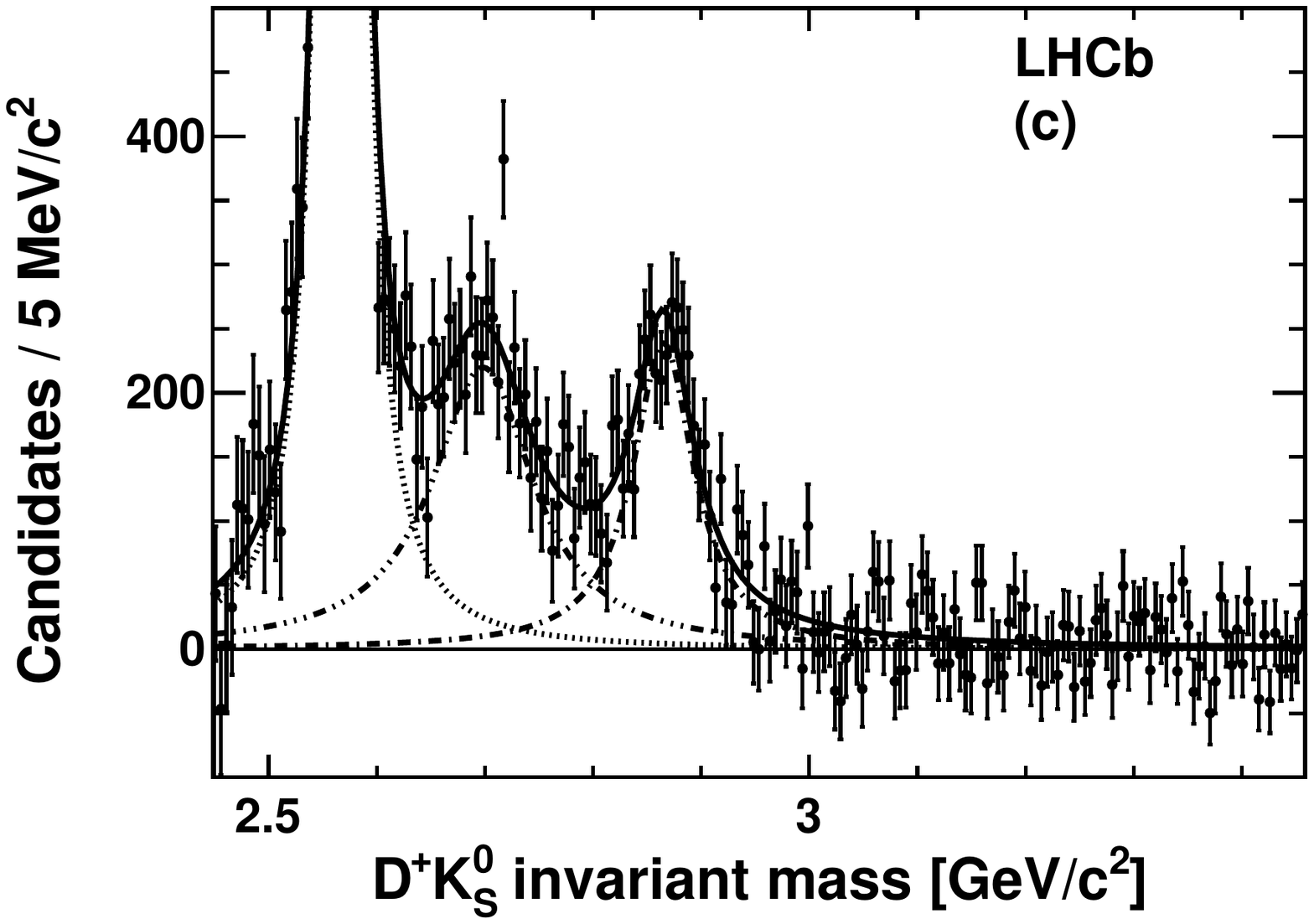}
  \includegraphics[width=0.49\textwidth]{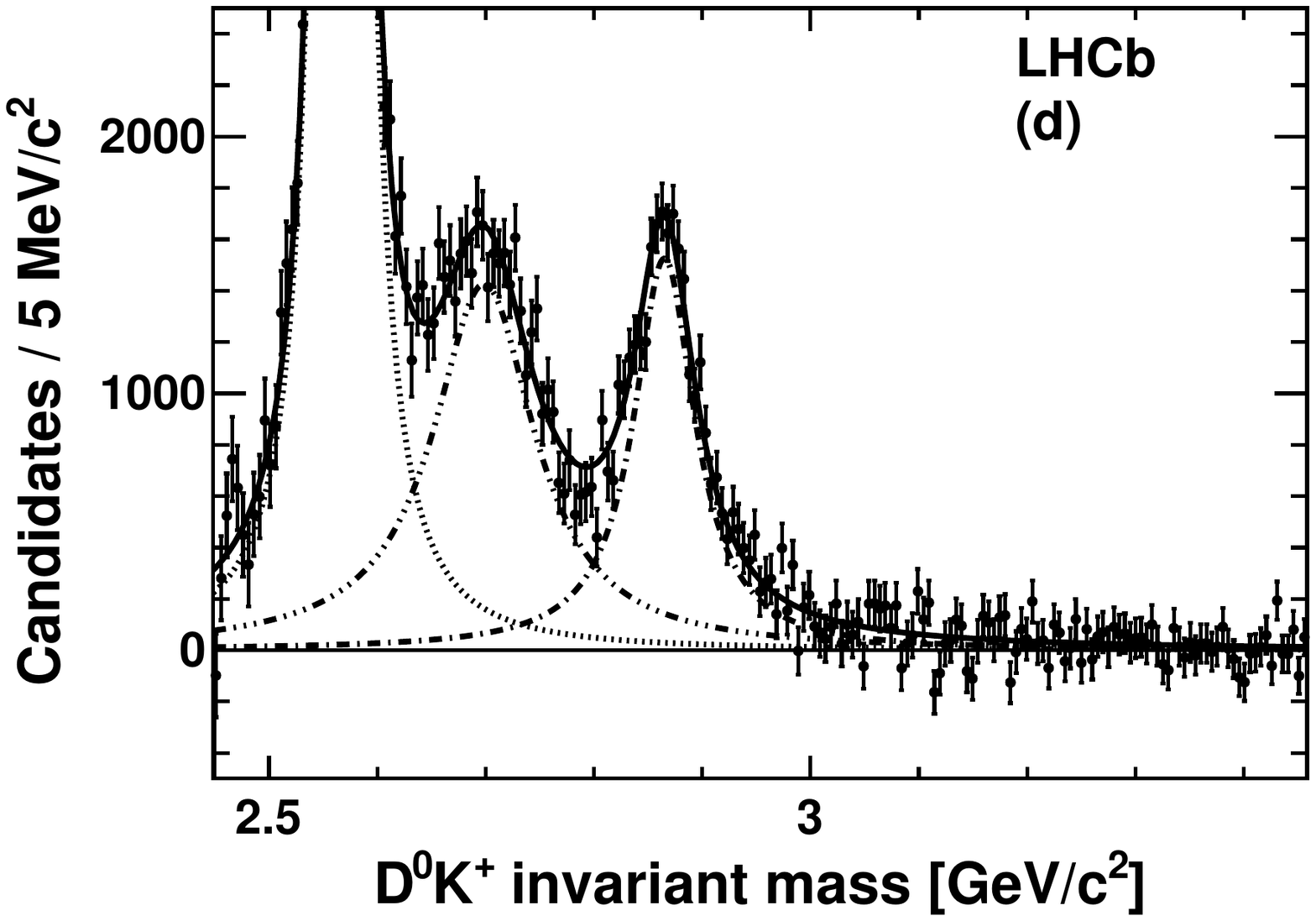}
    \vspace*{-0.5cm}
  \end{center}
  \caption{
    \small 
Invariant mass distributions (points) for (a) $D^+\KS$ and (b) $D^0K^+$. We show the total simultaneous probability density function (solid line), the $D_{s2}^*(2573)^+$ (fine dotted line), $D_{s1}^*(2700)^+$ (dot-dot-dot dashed line), $D_{sJ}^*(2860)^+$ (dot dashed line) and background contribution (dashed line). Invariant mass distributions after combinatorial background subtraction are shown for (c) $D^+\KS$ and (d) $D^0K^+$, where the vertical scales are truncated to show the $D_{s1}^*(2700)^+$ and $D_{sJ}^*(2860)^+$ signals more clearly.
    }
  \label{fig:fit}
\end{figure}

The reference fit includes a total of twenty-six parameters, fourteen to describe the background components (six fixed as mentioned above) and twelve for the description of the signal contributions. The six parameters for the masses and widths of all the $D_{sJ}$ structures are constrained to be the same in the $D^+\KS$ and $D^0K^+$ samples. The reference fit results for the $D_{s1}^*(2700)^+$ and $D_{sJ}^*(2860)^+$ parameters and total number of events are reported in Table~\ref{tab:fit} and Table~\ref{tab:fityields}, respectively. 
The projections of the fitted function superimposed to the data and the residuals after subtracting the fitted background distribution, are shown in Fig.~\ref{fig:fit}.

\begin{table}[tb]
  \caption{ \small
	Parameters for $D_{s1}^*(2700)^+$ and $D_{sJ}^*(2860)^+$ states, evaluated with binned fits to the samples. %
Masses and widths are given in units of \mevcc. Uncertainties are statistical only.}
\begin{center}\begin{tabular}{l|c|c|c|c|c}
 \multicolumn{2}{c|}{} & \multicolumn{2}{c|}{ $D_{s1}^*(2700)^+$ } & \multicolumn{2}{c}{ $D_{sJ}^*(2860)^+$ } \\
\hline
    Fit sample & $\chi^2$/ndf & $m$ & $\Gamma$   & $m$ & $\Gamma$        \\ 
    \hline
Reference fit to $D^+\KS$ and $D^0K^+$ & 464/422 & $2\,709\pm2$ & $115\pm7$  & $2\,866\pm1$ & $70\pm3$\\
$D^+\KS$ only fit          & 207/214 & $2\,710\pm4$ & $\,\,\,100\pm14$ & $2\,867\pm3$ & $73\pm7$\\
$D^0K^+$ only fit              & 241/214 & $2\,709\pm2$ & $117\pm8$  & $2\,866\pm1$ & $67\pm4$\\
\hline
  \end{tabular}\end{center}
\label{tab:fit}
\end{table}
\begin{table}[tb]
\caption{ \small Total number of events for $D_{s1}^*(2700)^+$ and $D_{sJ}^*(2860)^+$, evaluated with the reference fit. Uncertainties are statistical only.}
\begin{center}
\begin{tabular}{l|c|c}
Decay mode &  $D_{s1}^*(2700)^+$ & $D_{sJ}^*(2860)^+$  \\
\hline
$D^+\KS$ & $6\,724\pm596$ & $4\,825\pm347$ \\ 
$D^0K^+$   & $45\,315\pm2\,186$ & $31\,603\pm1\,257$\\
\hline
\end{tabular}
\end{center}
\label{tab:fityields}
\end{table}

The fit quality is acceptable with a total $\chi^2$/ndf of 464/422=1.1. 
We account for imperfections in the magnetic field map and alignment of the tracking system. These corrections are computed using a sample of $D^0\to K^-\pi^+$ decays, using the momentum scale calibration method explained in Ref.~\cite{Aaij:2011ep}. The corrections were found to be compatible with zero and therefore neglected.

\section{Cross-checks and systematic uncertainties}
\label{sec:Systematics}

The fit is validated using a large set of simulated experiments. 
No biases are observed and the resolution reported by the fit to data is found to be in agreement with the resolution from the analysis of the generated experiments. As a cross-check, we perform a set of fits to different data subsamples. We perform independent fits to the $D^+\KS$ and $D^0K^+$ samples (Table~\ref{tab:fit}) and to the $D^+\KS$ sample splitting the contributions from the $\KS$ meson decaying inside and outside the vertex detector. We repeat the reference fit on different $DK$ samples recorded with positive and negative magnet polarity, and also in a data sample of candidates required to pass dedicated $D^+$ and $D^0$ triggers. In all cases, we found the fit results to be compatible with the reference fit. 

\begin{table}[tb]
  \caption{ \small
	Systematic uncertainties for the $D_{s1}^*(2700)^+$ and $D_{sJ}^*(2860)^+$ parameters.  Mass and width uncertainties, $\delta m$ and $\delta\Gamma$, are given in units of \mevcc. The total uncertainties are calculated as the quadratic sums of all contributions.}
\begin{center}\begin{tabular}{l|c|c|c|c}
 & \multicolumn{2}{c|}{ $D_{s1}^*(2700)^+$ } & \multicolumn{2}{c}{ $D_{sJ}^*(2860)^+$ } \\
\hline
    Source &  $\delta m$ & $\delta\Gamma$   & $\delta m$ & $\delta\Gamma$       \\ 
    \hline
Signal model & 2.2 & 3.0 & 5.5 & 3.4 \\
Background model & 2.1 & 10.2 & 3.8 & 4.2 \\
High mass state & 0.0 & 0.3 & 0.0 & 0.2\\
Selection criteria & 2.1 & 3.5 & 1.0 & 2.7\\
Mass resolution & 2.1 & 3.6 & 2.8 & 2.4\\
Feed-down reflections & 1.2 & 2.9 & 0.1 & 1.4\\
Bin size & 0.2 & 0.9 & 0.0 & 0.2\\
\hline
Total & 4.5 & 12.1 & 6.3 & 6.6\\
\hline
  \end{tabular}\end{center}
\label{tab:syst}
\end{table}

Systematic uncertainties are summarized in Table~\ref{tab:syst}. There are calculated as the difference between the results of alternative fits and the reference fit, unless otherwise stated. 

A systematic uncertainty is associated to the signal model. Given the unknown $J^P$ assignment for the $D_{sJ}^*(2860)^+$ excited state, we repeat the reference fit assuming spin-1, spin-2 and spin-3 hypotheses for this resonance. A second systematic contribution to the signal description comes from the fact that the Blatt-Weisskopf form factors introduce a penetration radius that we fixed in the reference fit to $1.5\gev^{-1}$. The contribution to the systematic uncertainty is estimated by varying this value within the $1-3\gev^{-1}$ range. In both cases, we take the largest variation as systematic uncertainty. The quadratic combination of these two effects represents the largest systematic contribution to the $D_{sJ}^*(2860)^+$ parameters.

The background component is highly correlated with the yield and width of the broad structures, particularly for the $D_{s1}^*(2700)^+$ state. Four uncorrelated effects are studied. We use an empirical function to describe the background component in the $D^+\KS$ decay mode. This function, similar to that used in the \babar analysis~\cite{Aubert:2009ah}, is composed of a threshold function multiplied by a decreasing exponential of the form $\left(m-m_{\text{th}}\right)^p\exp\left\{-c_1m-c_2m^2\right\}$, where $m_{\text{th}}=m(D^+)+m(\KS)$. On the $D^0K^+$ sample, this function does not reproduce correctly the background shape.
Instead we generate a set of samples, using the reference probability density function, but randomly varying the background parameters. The average difference between the generated and fitted values for the $D_{s1}^*(2700)^+$ and $D_{sJ}^*(2860)^+$ masses and widths is taken as the systematic uncertainty. We repeat the reference fit changing the lower bound of the fit range by $\pm$10\mevcc and the upper bound by $-50\mevcc$. This has the largest effect on the width of the $D_{s1}^*(2700)^+$ state since the broad width is sensitive to modifications in the amount of background near the threshold and in the long high-mass tail.
Finally we evaluate a systematic uncertainty given by the effect of fixing some of the background parameters in the reference fit. 
We perform a set of fits accounting for all possible up and down variations (independently and simultaneously) of these parameters. The variations are of 10\% for  $D^+\KS$ background parameters and of 5\% in the case of the $D^0K^+$ decay mode. According to a fit $\chi^2$ study, alternative fits with larger variations of the fixed parameters do not describe the data correctly and therefore not used to compute systematic uncertainties. We adopt as systematic uncertainty the root-mean-square variation of all the fits for the given parameter. As expected, this effect contributes mainly to the widths of the resonances since these parameters correlate strongly with the background shape. The total background model systematic uncertainty is the quadratic combination of the four effects discussed.

Evidence for an additional broad state around 3\gevcc has been shown previously in $D^*K$ decay modes~\cite{Aubert:2009ah}. Therefore, in addition to the $D_{s1}^*(2700)^+$ and $D_{sJ}^*(2860)^+$ high mass states, we allow for another signal component in the fit.
No statistically significant structure is found.

The uncertainty introduced by the selection criteria is computed by repeating the fit in a sample with the following selection: $\pt(D^+\KS)>4.75\gevc$ and $\pt(\KS)>1.7\gevc$ for $D^+\KS$ combinations with the $\KS$ meson decaying inside and outside the vertex detector, respectively, while for the $D^0K^+$ sample we apply $\pt(K^+)>1.8\gevc$ and $P_{\text{NN}_K}(K^+)>0.5$. These selection criteria are established by optimizing the signal significance of the $D_{s2}^*(2573)^+$ in the $2.5-2.6$\gevcc range, as done previously, but this time downscaling the number of signal events by one order of magnitude $0.1N_S/\sqrt{0.1N_S+N_B}$, trying to mimic the signal to background ratio observed for the $D_{s1}^*(2700)^+$ and $D_{sJ}^*(2860)$ states.

Mass resolution effects are neglected in the reference fit since the measured widths are much larger than the mass resolution obtained from Monte Carlo simulated data: 4.3\,\,(3.3)\mevcc at 2.71\gevcc and 5.2\,\,(4.0)\mevcc at 2.86\gevcc mass for the $D^+\KS(D^0K^+)$ decay mode. This effect is accounted for by a convolution of the relativistic Breit-Wigner lineshapes with a single Gaussian function without offset whose width is fixed to the mass resolution estimated using fully simulated events. Here, the largest contribution arises from the $D_{s2}^*(2573)^+$ state, since a narrower width for this state causes a deviation in the masses and widths of the resonances under study. 

The observed $D_{s1}^*(2700)^+$ and $D_{sJ}^*(2860)^+$ states can also decay into $D^*K$ final states (depending on the $D_{sJ}^*(2860)^+$ spin-parity) and this should be reflected as feed-down components to the $DK$ samples, arising from $D^{*+}\to D^+\pi^0,\,D^+\gamma$ and $D^{*0}\to D^0\pi^0,\,D^0\gamma$ decays, where the neutral pion and photon are not reconstructed. In this case, we expect the feed-down structures to be shifted by about $-142\mevcc$ from the measured mass and with similar width but with a small spread from resolution effects. Ignoring resolution effects, we evaluate a systematic uncertainty due to the presence of possible feed-down by including the two additional components to describe the $D_{s1}^*(2700)^+\to D^{*+}\KS,\, D^{*0}K^+$ and $D_{sJ}^*(2860)^+\to D^{*+}\KS,\, D^{*0}K^+$ processes, with fixed masses and widths to avoid large correlations. The uncertainty due to this effect is about a factor two smaller than the statistical precision on the masses and widths.  

Finally, to investigate the effect of binning the data samples, we repeat the fit using bins with size of 1\mevcc. This effect is observed to be negligible.

The total systematic uncertainty is calculated as the quadratic sum of all the mentioned contributions. The systematic uncertainties on the $D_{s1}^*(2700)^+$ and $D_{sJ}^*(2860)^+$ parameters dominate the overall measurement uncertainties.

\section{Conclusions}
\label{sec:Conclusions}
Using 1.0\invfb of data recorded by the \lhcb experiment during 2011 in $pp$ collisions at a centre-of-mass energy of $\sqrt{s}=7\tev$, we perform a study of the $D^+\KS$ and $D^0K^+$ final states. We observe for the first time the production of $D_{s1}^*(2700)^+$ and $D_{sJ}^*(2860)$ states in hadronic interactions and measure their parameters to be 
\begin{eqnarray}
m(D_{s1}^*(2700)^+) &=& 2709.2 \pm 1.9(\mbox{stat})\pm\,\,\,4.5(\mbox{syst})~\mevcc,\cr
\Gamma(D_{s1}^*(2700)^+) &=& \,\,\,115.8 \pm 7.3(\mbox{stat}) \pm12.1(\mbox{syst})~\mevcc,\cr
m(D_{sJ}^*(2860)^+) &=& 2866.1 \pm 1.0(\mbox{stat}) \pm\,\,\,6.3(\mbox{syst})~\mevcc,\cr
\Gamma(D_{sJ}^*(2860)^+) &=& \,\,\,\,\,\,69.9 \pm 3.2(\mbox{stat}) \pm\,\,\,6.6(\mbox{syst})~\mevcc.\cr\nonumber
\end{eqnarray}
All results are compatible with previous results from the $B$ factories~\cite{Brodzicka:2007aa,Aubert:2009ah}. The statistical uncertainties for all parameters are improved by an overall factor of two with respect to the \babar measurements in the same decay modes, and it is of the same order as for the combined $DK$ and $D^*K$ \babar measurement. The precision of the measured quantities is dominated by systematic effects. We do not observe any statistically significant $D_{sJ}$ resonance in the mass region above 3\gevcc.

To shed light on the puzzle around the spin-parity of the $D_{sJ}^*(2860)^+$ state and to confirm the spin-parity assignment of the $D_{s1}^*(2700)^+$, an angular analysis of $D^{*}K$ samples would be needed.

\section*{Acknowledgements}

\noindent We express our gratitude to our colleagues in the CERN accelerator
departments for the excellent performance of the LHC. We thank the
technical and administrative staff at CERN and at the LHCb institutes,
and acknowledge support from the National Agencies: CAPES, CNPq,
FAPERJ and FINEP (Brazil); CERN; NSFC (China); CNRS/IN2P3 (France);
BMBF, DFG, HGF and MPG (Germany); SFI (Ireland); INFN (Italy); FOM and
NWO (The Netherlands); SCSR (Poland); ANCS (Romania); MinES of Russia and
Rosatom (Russia); MICINN, XuntaGal and GENCAT (Spain); SNSF and SER
(Switzerland); NAS Ukraine (Ukraine); STFC (United Kingdom); NSF
(USA). We also acknowledge the support received from the ERC under FP7
and the Region Auvergne.

\addcontentsline{toc}{section}{References}
\bibliographystyle{LHCb}
\bibliography{main}

\end{document}